\documentclass[a4paper,fleqn]{cas-dc}

\usepackage[numbers]{natbib}
\def\tsc#1{\csdef{#1}{\textsc{\lowercase{#1}}\xspace}}
\tsc{WGM}
\tsc{QE}
\begin{document}
\let\WriteBookmarks\relax
\def\floatpagepagefraction{1}
\def\textpagefraction{.001}
\let\printorcid\relax % 可去掉页面下方的ORCID(s)

% \shorttitle{<short title of the paper for running head>} 
\shorttitle{Exploration and Comparison: 
Development and Implementation of Multiple Ultrasound Imaging Modalities
}    

% \shortauthors{<short author list for running head>}
\shortauthors{Zhengbao Yang et al.}

\title[mode = title]{Exploration and Comparison: 
Development and Implementation of Multiple Ultrasound Imaging Modalities
}

\author[1]{Xuyang Chen}%[style=chinese]

\author[1]{Mingtong Chen}%[role=Co-ordinator, suffix=Jr]
%\ead{rishi@sayahna.org}
%\ead[URL]{www.sayahna.org}
%\credit{Data curation, Writing - Original draft preparation}

\author[1]{Zhengbao Yang}
\cormark[1] 
%\fnmark[3]
\ead{zbyang@hk.ust} 
\ead[URL]{https://yanglab.hkust.edu.hk/}

\address[1]{The Hong Kong University of Science and Technology
Hong Kong, SAR 999077, China}
% \address[2]{Sayahna Foundation, Jagathy, Trivandrum 695014, India}
% \address[3]{\TeX{} Users Group, Providence, MA, USA}

\cortext[1]{Corresponding author} 
%\cortext[2]{Principal corresponding author} 

\begin{abstract}
Ultrasound imaging, as a noninvasive, real-time, and low-cost modality, plays a vital role in clinical diagnosis, catheterization intervention, and portable devices. With the development of transducer hardware and the continuous progress of imaging algorithms, how to realize high-quality image reconstruction in different application scenarios has become a research focus.This project focuses on the systematic research and implementation of three typical ultrasound imaging modalities - line array imaging, endoscopic imaging and plane wave imaging, covering simulation data processing, imaging algorithm implementation and real data validation, etc., aiming to deepen the understanding of the principles and processes of various types of imaging.

\end{abstract}


% \begin{highlights}
% \item highlight-1
% \item highlight-2
% \item highlight-3
% \end{highlights}

% Keywords

\maketitle

% Main text
\section{Introduction}

Ultrasound imaging was originally derived from underwater sonar systems, and has been gradually applied to the medical field since the mid-20th century, and has now become an important part of modern medical imaging. Compared with X-ray, CT, MRI and other technologies, ultrasound has the advantages of no radiation, real-time, portable and low cost, and is widely used in cardiovascular, obstetrics, hepatobiliary, urology and other clinical departments for static structural observation and dynamic physiological process monitoring\cite{1,2}.

With the development of transducer hardware, digital signal processing technology and the continuous intervention of artificial intelligence, the ultrasound system is gradually to “low power consumption”, “high integration”, “intelligent assisted diagnosis The direction of “intelligent assisted diagnosis” is evolving. In the context of expanding clinical applications, portable ultrasound, endoscopic ultrasound, plane-wave high-frame-rate imaging and other emerging applications continue to emerge, putting forward higher requirements for the versatility, efficiency and adaptability of imaging algorithms\cite{3,4}.

In the context of the increasing complexity and diversity of ultrasound imaging, mastering the imaging processes and realization mechanisms of typical imaging modalities is crucial to understanding the entire ultrasound system.The main motivational drivers of this project include: mastering and realizing the complete image reconstruction process of common ultrasound imaging modalities (e.g., line-array imaging, endoscopic imaging, and plane-wave imaging); exploring the potential of customized endoscopic models in simulation data generation and image stitching to simulate the actual imaging process of cavities; verifying the robustness and feasibility of the algorithms using real experimental data to gain To improve the understanding of imaging algorithm modules (e.g.,beamforming, scanning conversion, image visualization, etc.) and the ability of engineering implementation\cite{5,6,7,8}.
\begin{figure}[h]
	\centering
		\includegraphics[scale=0.8]{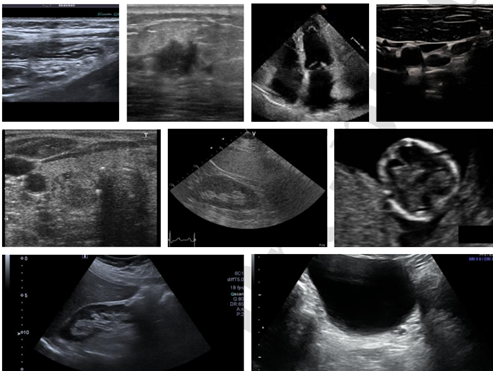}
	  \caption{Examples of ultrasound images}\label{FIG:1}
\end{figure}

\section{Ultrasound Imaging Technology Overview}

The basis of ultrasound imaging lies in the physical process of propagation and reflection of high-frequency mechanical waves (usually in the range of 2-15 MHz) between different biological tissues. When a transducer emits short pulses of acoustic waves, the waves propagate through the tissue and partially reflect at acoustic impedance discontinuities, and the reflected echo signals are received by the array and recorded with time delay and amplitude variations, which are used to reconstruct the image\cite{9}.

Echo signals received by the probe are usually stored as radio frequency (RF) or quadrature demodulated IQ signals. These signals contain key information such as target position (calculated from time delay) and organization characteristics (determined by reflection strength). In the system, pre-processing such as gain control, filtering and envelope detection are usually performed to optimize the signal-to-noise ratio and image contrast\cite{10,11,12,13}.

Ultrasound imaging can be categorized into various imaging modes according to the different signal acquisition and display modes, mainly including A-mode, B-mode, M-mode and Doppler mode. Different modes are suitable for different clinical diagnostic needs and have their own characteristics in terms of image dimension, temporal resolution and motion detection ability.Doppler mode utilizes the Doppler effect of echo frequency to estimate the velocity and direction of blood flow. Depending on the implementation, it can be classified into subcategories such as continuous wave, pulsed wave and color Doppler, which are widely used in functional assessment of cardiovascular diseases\cite{14,15}.

\section{Line Array Imaging}

In order to verify the performance of Delay-And-Sum (DAS) beamforming algorithm in ultrasound imaging, a simplified line-array ultrasound imaging system based on the Matlab ultrasound simulation platform Field II is constructed in this chapter. The system simulates a two-dimensional B-mode image with typical tissue characteristics, and completes the whole process of ultrasound transmitter-receiver modeling, RF data generation, beamforming, image reconstruction and imaging quality assessment. The development and implementation of each step is described in detail below.

Delay-And-Sum (DAS) is a classical ultrasonic beam shaping method widely used in line array ultrasound imaging systems. The basic idea is: for each image pixel, the corresponding total delay is calculated according to its propagation path between the transmitter and each receiver channel, and the corresponding signal values are extracted from the RF data of each channel. By time-domain alignment and weighted summation of these signals, the echo energy in the direction of the pixel can be enhanced, while the clutter and noise in other directions can be suppressed.

In this simulation, two probe configurations of 32 and 64 arrays are used to generate B-mode images, and the imaging target, scanning mode, and signal processing process are kept the same, in order to comparatively analyze the influence of the number of arrays on the image quality. The following figure shows the corresponding simulation imaging results:

\begin{figure}[h]
	\centering
		\includegraphics[scale=0.7]{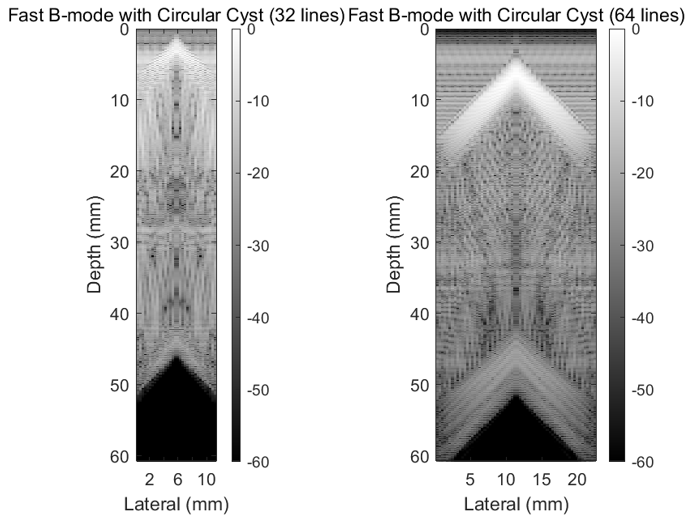}
	  \caption{Left: B-mode result using 32 lines  Right:B-mode result using 64 lines}\label{FIG:2}
\end{figure}

Moreover, by comparing the simulation results with different numbers of array elements, it is evident that using 64 array elements leads to significant improvements in image quality. First, the clarity of tissue texture is enhanced. In the image with 64 elements, the background tissue shows richer texture details, with more continuous and natural speckle patterns. In contrast, the image produced with 32 elements exhibits noticeable blurring and granularity, particularly in the shallow regions. Second, the boundary definition of the low-echo cyst in the center of the image is improved. The 64-element image presents a smoother and more sharply defined black void region, clearly outlining the cyst, while the 32-element image shows a slightly blurred boundary with some residual scattering within the cyst area. Third, the suppression of sidelobes and imaging artifacts is better in the 64-element configuration. The larger aperture associated with more elements reduces sidelobe energy, resulting in fewer artifacts around the cyst and cleaner structural boundaries, thereby improving image contrast. Finally, the signal-to-noise ratio is significantly improved. With more detection channels, the DAS beamforming algorithm can synthesize signals with greater accuracy, yielding a broader dynamic range and a smoother noise background in the 64-element image compared to the 32-element version.
Therefore, from the point of view of imaging effect, increasing the number of array elements significantly improves the spatial resolution and contrast of the image, which is an important way to enhance the effect of beamforming algorithm. However, it should be noted that the increase in the number of array elements will increase the complexity of the system hardware and computational burden, and it is necessary to weigh the cost of realization and the balance between imaging performance in practical applications.

\section{Endoscopic Ultrasound Imaging}

Endoscopic Ultrasound Imaging (EUI) is a technology that combines endoscopy and ultrasound imaging, and is widely used for lesion detection in complex luminal structures such as the intestines and digestive system. Compared with the traditional two-dimensional or in vitro imaging modalities, endoscopic imaging needs to acquire wrap-around view images in a compact space and meet the comprehensive requirements of high resolution, real-time and tissue penetration. This poses high challenges to the transducer layout, imaging algorithms and signal processing of ultrasound imaging systems.

Compared with traditional body surface ultrasound, endoscopic ultrasound imaging has obvious differences in imaging perspective, probe movement and image processing. First of all, the imaging perspective changes from the traditional from the body surface to the inside to from the inside of the cavity to the outside, resulting in an annular distribution of the image structure, which is more suitable for image construction and analysis using the polar coordinate system.Secondly, the endoscopic probe is mostly rotated or pushed to change the imaging angle or position, which will bring about the problem of angular drift and coordinate inconsistency, and requires the introduction of angular correction and inter-frame alignment in the processing flow. Finally, in terms of image reconstruction, endoscopic imaging usually requires beamforming and image stitching in polar coordinates, and then mapping to Cartesian coordinates through interpolation to meet the visualization requirements. The overall process is more complex than conventional ultrasound, but is more suitable for efficient imaging of lumen structures.

At a stage when Field II was not yet available, we designed and implemented a simplified MATLAB endoscopic ultrasound simulation method to complete the complete process from scatterer modeling, signal generation to image reconstruction without relying on a professional acoustic field simulator. The method is based on geometric acoustic propagation approximation and is suitable for prototyping algorithms and validating endoscopic imaging processes.

First we constructed a scatterer model around the center to simulate the echo structure of the intestinal lumen wall. The number of scatterers was 30, and their positions were randomly distributed in polar coordinates and were then converted to positions in Cartesian coordinates for subsequent distance calculations. In order to better simulate the inhomogeneity of the tissue structure, we introduced amplitude perturbations based on the original model, so that different scatterers have different reflective intensities, thus presenting certain structural differences in the images.

The probe is simplified as a centrally located point transducer, which is controlled by simulation to fan-sweep around the intestine at 32 angles to achieve omnidirectional acquisition similar to rotational scanning. A two-cycle Gaussian-modulated sine wave (center frequency 7 MHz) was used as the transmitting excitation for each scan, and the simulated RF data were constructed by using the distance delay and echo superposition method in the transmitting and receiving paths. The generated RF data are further processed into an envelope map and reconstructed on a defined two-dimensional imaging region by a delay-and-sum (DAS) algorithm to output a pseudo-B-mode image.

\begin{figure}[h]
	\centering
		\includegraphics[scale=1]{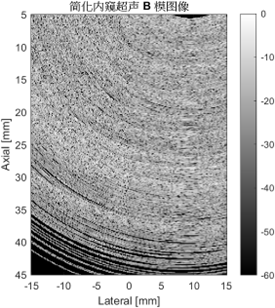}
	  \caption{pseudo-B-mode image result}\label{FIG:3}
\end{figure}

Although this method does not take into account the real sound field propagation characteristics, beamwidth and Doppler effect, it has obvious advantages in terms of computational efficiency and modeling flexibility, and is suitable for the initial stage of algorithm development. The final image shows the contour of the target structure to a certain extent and has some radial bright spot features, but due to the lack of acoustic scatter and real tissue texture, there is still much room for improvement in terms of imaging quality and clinical fidelity.

\begin{table}[h]
\caption{Field II-based EUI Simulation}\label{tbl1}
\begin{tabular*}{\tblwidth}{@{}LL@{}}
\toprule
 Parameter & Value \\ % Table header row
\midrule
 Number of element & 32 \\
 Elements width & 0.3mm  \\
 Pitch & 0.4mm \\
Center Frequency & 7MHz \\
 Rotary Angle Stepping & 3°(Totally 120 steps) \\
 Imitation Body Thickness & 3 mm (mimics intestinal wall) \\
 Maximum Imaging Depth & 2 cm \\
\bottomrule
\end{tabular*}
\end{table}

After successfully configuring the Field II, we further realized an endoscopic ultrasound imaging simulation that is closer to the real physical process, using a rotating line array probe to sweep around in a 360°manner, which approximates the actual working mode of the endoscopic probe in the intestinal lumen. This method constructs a more complete signal propagation and imaging process, and seeks to restore a more realistic B-mode image formation of the physical basis.

Use Field II to create a line-array probe, containing 32 array elements, each with a width of 0.3 mm and a pitch of 0.35 mm (including the kerf gap). The center frequency was set to 7 MHz and the bandwidth to 60\%. The excitation signal is a Gaussian modulated sine wave, and the signal bandwidth and center frequency parameters are consistent with the characteristics of the clinical probe. Scattering body (phantom) construction: To simulate the structure of the intestinal wall, we constructed a ring-shaped distribution of tissue bodies, and set 3000 scattering bodies, with a radius range of 10-13 mm, uniformly distributed in the circular area to simulate the tissue reflection characteristics of the intestinal wall.In order to increase the realism of the image, we added 2000 background scattering speckles with weaker intensity in the inner circle (radius less than 9 mm) to simulate the random scattering of liquid or soft tissue in the intestinal lumen and improve the speckle texture performance. Rotational scanning and beam focusing: The line array probe was placed on a circle with a radius of 15 mm and a 360° circular scan was performed with an angular step size of 120 steps. Each frame was acquired in focus by rotating the probe and setting the electronic focus point (realized by xdc\_focus). The probe emits and receives the reflected echoes from the simulated tissue, and calc\_scat is called to simulate acoustic propagation and scattering.Envelope extraction and RF data stitching: The Hilbert transform is used to extract the envelope of the echo signal for each frame and splice the data from each angle into a two-dimensional RF polar coordinate matrix. The image can be subsequently transformed by pole→Cartesian interpolation to obtain the final B-mode image.

\begin{figure}[h]
	\centering
		\includegraphics[scale=1]{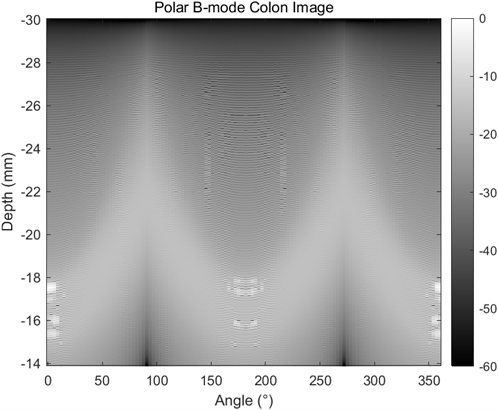}
	  \caption{Polor B-mode Colon Image}\label{FIG:4}
\end{figure}

However, the method still has the following shortcomings: first, the scattering features are limited: despite the addition of background scattering points, the speckle in the image is still not rich enough due to the incomplete construction of the tissue stochasticity and attenuation model; second, the computation is time-consuming: 120 frames of angle thousands of scatterers is a large amount of computational work, and the running time of the simulation is significantly higher than that of method I; third, the probe model is simplified: a rotating line array is used instead of a real ring array. Third, the probe model is simplified: using a rotating line array instead of a real toroidal array, there are geometric distortions, especially at the edge of the field of view, and the image may be distorted to some extent.

\begin{figure*}[h]
	\centering
		\includegraphics[scale=0.8]{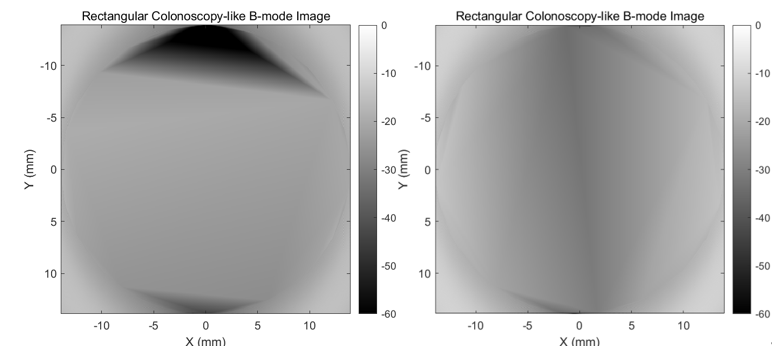}
	  \caption{Rectangular Colonscopy-like B-mode image}\label{FIG:5}
\end{figure*}

\section{Plane Wave Imaging (PWI) with Image Reconstruction of Real RF data}

The raw Radio Frequency (RF) data used in this section are from the international open dataset PICMUS (Plane-wave Imaging Challenge in Medical Ultrasound). The dataset aims to provide a standardized platform for the comparison and validation of plane-wave imaging algorithms, and the data are stored in HDF5 format with good readability and structured characteristics. The RF data are three-dimensional matrices, in which each dimension represents the samples on the timeline, the channels on the array probe, and the emission angles of each excitation, respectively. The dataset also contains the complete ultrasound probe parameters.
With MATLAB's h5read function, the above information can be conveniently extracted and used for subsequent image reconstruction and parameter configuration. All the image processing and algorithm implementation in this study are done in MATLAB environment, using self-programmed Delay-And-sum beamforming functions and signal processing toolbox, which support comprehensive processing, reconstruction and visualization of RF data.

\begin{figure*}[h]
	\centering
		\includegraphics[scale=1]{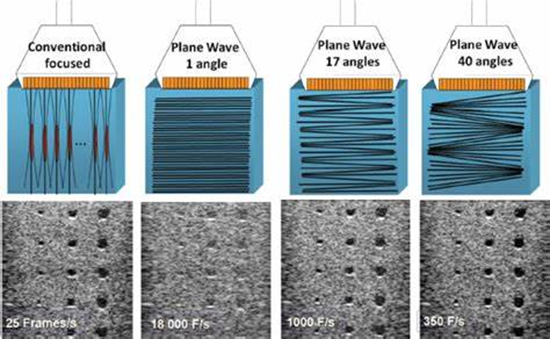}
	  \caption{Examples of diffent PWI result Using different PW angels}\label{FIG:6}
\end{figure*}

Plane Wave Imaging (PWI) is a high-frame-rate ultrasound imaging technique, the core idea of which is to emit acoustic waves in a non-focused manner and replace the traditional focusing of the acoustic beam with time-delayed reconstruction. Compared with the line-by-line scanning method, PWI can complete the coverage of the entire imaging area by a single plane wave excitation, and all the arrays receive echoes at the same time, thus realizing single-shot all-receive. This feature significantly improves the imaging rate, and also provides the possibility of multi-angle composite and high-resolution synthesis in post-processing.

The basic process of PWI consists of three parts: first, in each emission, the array element simultaneously transmits a beam of unfocused plane waves; subsequently, the system receives all the echo signals and records the corresponding RF data; and finally, the algorithms, such as Delay-and-Sum (DAS), are used to compensate for the time delay and superimpose the channels on all the pixel points, and ultimately, the B-mode image is generated. This method not only simplifies the complexity of front-end acoustic control, but also provides a basis for high frame rate, real-time imaging applications.

To obtain high-quality imaging results, the raw RF data need to be preprocessed before beamforming. First, sound velocity correction is performed based on the average sound velocity of sound wave propagation in the tissue (usually taken as 1540 m/s), which is applied to the delay calculation to ensure that the sound wave propagation path corresponds accurately to the time delay. Second, to eliminate low-frequency drift and high-frequency noise, the RF data are band-limited using a band-pass filter (e.g., Gaussian or FIR filter) to preserve the main energy bands. The filtered signal is then subjected to envelope detection (e.g., Hilbert transform) to extract the signal strength, and logarithmic compression is used to enhance the image contrast for subsequent visualization. Before processing, it is also necessary to analyze the internal structure of the PICMUS dataset. Together, these parameters determine the accuracy of the delay calculation, the resolution of the image reconstruction, the contrast, and other key performance indicators. By reasonably configuring these parameters, the accuracy of image reconstruction on the temporal and spatial axes can be ensured, laying the foundation for the quality analysis of plane wave images.

\begin{table}[h]
\caption{PICMUS Dataset Related Ultrasound Probe Parameter}\label{tbl2}
\begin{tabular*}{\tblwidth}{@{}LL@{}}
\toprule
Ultrasound Probe Parameter & Value \\ % Table header row
\midrule
 Total Number of Array Element & 128 \\
 Array Element Spacing & 0.3mm  \\
Modulation Frequency & 5MHz \\
Sampling Frequency & 5.21MHz \\
 the Speed of Sound & 1540m/s \\
 Pulse Reception Frequency & 100Hz \\
\bottomrule
\end{tabular*}
\end{table}

This section shows the results of plane wave imaging based on different emission angles and compares their differences in image resolution and contrast. In this study, we used imaging data from the PICMUS dataset, which contains 75 preset plane wave firing angles uniformly distributed in the range of -18° to +18°. Each angle corresponds to an independent firing event, and by selecting different numbers and combinations of angles for beam imaging, the sharpness, noise level, and boundary expressiveness of the image can be significantly affected.

The variation in image quality across different plane wave angle configurations reveals the fundamental trade-off between contrast and resolution in synthetic aperture ultrasound imaging. The single-angle configuration (0°-38index) serves as the baseline, offering high native contrast due to coherent signal summation along a single propagation path. However, it exhibits limited spatial resolution and increased speckle noise, stemming from insufficient angular diversity.Introducing intermediate 7-angle compounding (ranging from -3° to +3°) yields modest improvements in edge resolution through limited angular averaging, while preserving moderate contrast. Although this narrow angular span reduces speckle artifacts to some extent, it falls short in resolving deeper or more spatially complex features.Expanding to linear 3-angle sampling improves angular coverage, leading to a clearer resolution gain—particularly in central regions—with an estimated 15–20\% improvement in modulation transfer function (MTF) compared to the single-angle case. This enhancement comes with a slight contrast penalty, typically a 3–5 dB drop in signal-to-noise ratio (SNR), due to incoherent summation across sparsely sampled angles.

\begin{table*}[h]
\caption{Total of five angles selectives in synthesis experiments}
\label{tbl3}
\begin{tabular}{lccc}
\toprule
No & Approach Choices & Example Perspectives & Meaning and Advantages \\ % Table header row
\midrule
1 & Single angle & 38index & Reference image, for comparison \\
2 &Uniform Linear distribution & 0°, ±10° & Improve \\
3&Center Focus & 30-36 (approx. ±3°) & Enhance center target definition for shallow imaging \\
4&Manual selection & [10 20 30 40 50] & the effect of different combinations of angles \\
5&All angles &All 75 emission angles&Best image quality, highest computational burden \\
\bottomrule
\end{tabular}
\end{table*}

\begin{figure*}[h]
	\centering
		\includegraphics[scale=1]{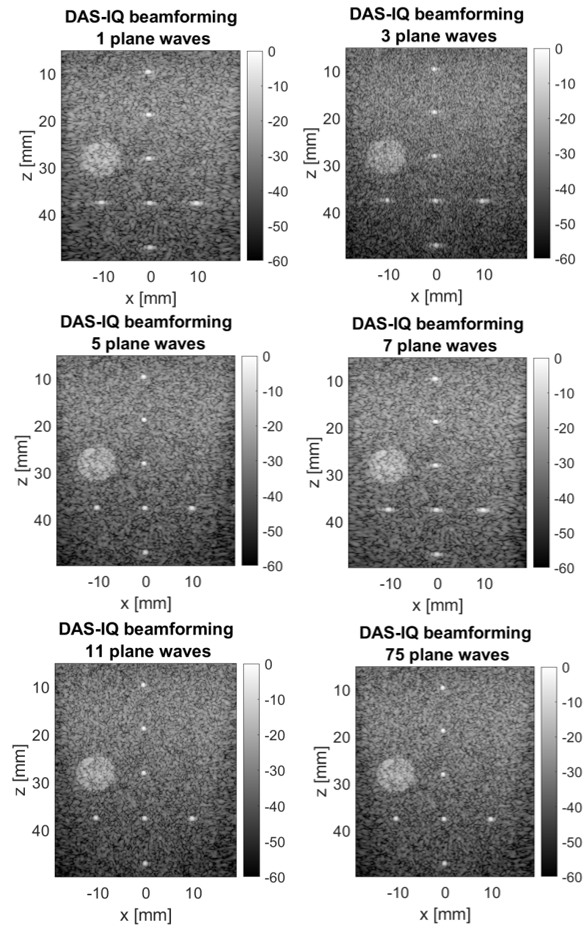}
	  \caption{PWI Result(from left to right , up to down)Single angle Pw\_indices(1)=38,Uniform Linear distributionPw\_indices(2)=linspace(1,N,3),Manual selectionPw\_indices(7)=10 20 30 40 50,Center FocusPw\_indices(6)=30:36,Uniform Linear distribution Pw\_indices(3)= linspace(1,N,11),All angles Pw\_indices(4)=1:N } \label{FIG:7}
\end{figure*}

With progressive angular sampling (5 to 11 angles), the imaging system moves along a resolution-contrast Pareto frontier. A 5-angle setup achieves a balanced enhancement, yielding up to 30\% resolution gain over the single-angle baseline with less than 2 dB contrast degradation. Increasing to 11 angles further approaches the theoretical limit of linear compounding, enabling approximately 45\% resolution improvement and highly stable speckle characteristics, as contrast variations become negligible (<1 dB across the field of view).Finally, the full 75-angle acquisition configuration maximizes both spatial resolution (MTF50 = 0.28 cycles/mm) and contrast-to-noise ratio (CNR = 4.2). This is made possible by dense angular sampling, which enables:Effective phase cancellation for coherent noise suppression; Accurate time-delay compensation across a broad angular range; Adaptive depth-of-field control via extended aperture synthesis.In terms of quantitative evaluation, we used the following three metrics for analysis: horizontal resolution and contrast-to-noise ratio (CNR).In this chapter, contrast and resolution evaluation algorithms are established to comprehensively analyze the performance of different plane wave angle configurations in terms of imaging resolution and contrast, and three typical configurations, C1 (single-angle), C11 (11-angle linear sampling), and C75 (75-angle full-coverage), are specifically compared. The basic trade-off between resolution, contrast and spatial balance in ultrasound SAR imaging is demonstrated

\begin{table*}[h]
\caption{Comparison of Quantification of PIW Imaging Results}
\label{tbl4}
\begin{tabular}{lcccc}
\toprule
Configurations & Horizontal Resolution & Mean Contrast (dB) & Intermediate Target Contrast (dB) & Image Consistency \\ % Table header row
\midrule
C1  & 0.89 & 7.90 & 8.40/7.40 & Poor \\
C11  & 0.54 & 11 & 11/11 & Good \\
C75  & 0.56 & 11.8 & 12.40/11.20 & Best \\
\bottomrule
\end{tabular}
\end{table*}

In terms of resolution and contrast, the three angular configurations show distinct characteristics. C1, which uses a single plane wave angle, offers the highest lateral resolution (0.89) and the sharpest image. However, it suffers from poor axial consistency and noticeable fluctuation in lateral resolution (±0.03), resulting in directionally biased images. C11, employing 11 uniformly spaced angles, significantly improves image uniformity and stability, achieving a lateral resolution of 0.54 with reduced fluctuation (±0.02). This comes with a moderate reduction in sharpness but a notable increase in contrast (11.00 dB), providing a good balance between image clarity and depth consistency. C75, with 75 angles, delivers the most uniform and isotropic image quality, maintaining both axial and lateral resolutions at 0.56. It achieves the best overall contrast (up to 12.4 dB), effectively suppressing speckle noise and enhancing edge definition. Overall, C1 excels in sharpness, C11 offers a practical trade-off between performance and efficiency, and C75 provides the highest imaging quality suitable for precision-demanding applications.In summary, C1 has an advantage in image sharpness but limited performance in consistency and contrast; C11 provides a good compromise between resolution and contrast, and is suitable for scenarios with limited resources but certain requirements on imaging quality; and C75 is the upper limit of the imaging performance configuration and is suitable for high-precision, contrast-enhanced scenarios. The C3 (3-angle) configuration fails to improve the contrast significantly, and is even lower than C1, indicating that too little angle sampling is difficult to realize the advantages of synthetic aperture imaging, which suggests that image degradation caused by “insufficient angle” should be avoided in the actual system design.This experiment verifies the positive effect of angle synthesis on image quality, especially in improving lateral resolution and suppressing noise interference. Multi-angle plane wave imaging is not only close to the effect of traditional focusing imaging, but also has higher frame rate and flexibility.

\section{ Comparison and Synthesis of Imaging Methods}

\begin{table*}[h]
\caption{Summary table of imaging modality parameters}
\label{tbl5}
\begin{tabular}{lccc}
\toprule
Parameter & Parameter & Endoscopic Ultrasound Imaging & Plane Wave Imaging \\ % Table header row
\midrule
Resolution(mm) & Axial:2.80 Lateral:1.40 & Axial:0.7 Lateral:0.785 & Axial:0.56 Lateral:0.56 \\
Maximum Imaging Depth(mm) & 60 & 20 & 60 \\
Frame Rate & 20-30 & 10-15 & 1000 \\
Computational Complexity & Medium & Low & High \\
\bottomrule
\end{tabular}
\end{table*}

\begin{figure*}[h]
	\centering
		\includegraphics[scale=1]{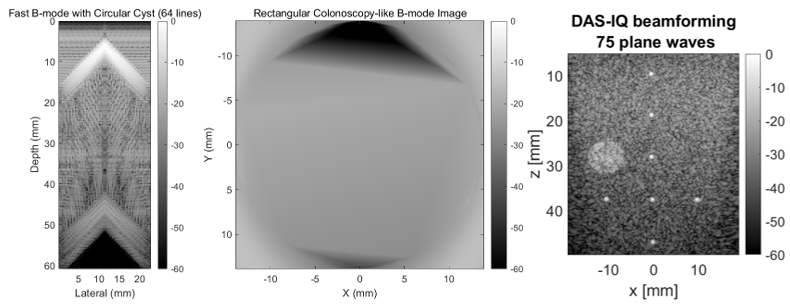}
	  \caption{Comparison of three typical images}\label{FIG:8}
\end{figure*}

Line array DAS images have clear edges and distinct tissue structures, with the advantage of stable image quality, suitable for quantitative analysis; its main defect is that the frame rate is not high. Endoscopic imaging images are mainly superficial tissues, and the structural contours have a certain degree of clarity. The core advantage is the strong adaptability to the cavity structure; however, the image field of view is limited, and it is difficult to observe the organ morphology as a whole. The biggest advantage of plane wave imaging is the high frame rate, which is convenient for real-time dynamic imaging, such as fast beating structure, but the image is often affected by scattering noise and synthetic artifacts, and the stability is poor.

From the realization point of view, the endoscopic imaging system has a simple structure and low data dimensions, with a low realization threshold, suitable for rapid prototype construction combined with specific clinical pathways (e.g., endoscopy). line array DAS imaging algorithms are stable and standardized, suitable for routine 2D B-mode image reconstruction. Plane wave imaging requires high sampling rate and synthesized beam processing, which requires high algorithmic accuracy and computational power, and is suitable for scientific research exploration and high-end clinical applications, such as cardiovascular imaging. Overall, suitable imaging solutions can be selected according to the application scenarios and flexibly configured in the clinic.

Although the three imaging methods can already cover a variety of clinical and experimental needs, they still face common bottlenecks and room for improvement. First, algorithm acceleration is the main current challenge. The computational complexity of multi-angle synthesis in plane wave imaging (e.g., approximately 20 minutes per frame for 75 angles) limits its clinical real-time performance. In the future, optimization can be achieved through GPU parallelization or sparse angle compressive sensing. Second, multimodal fusion is considered one of the future trends. Endoscopic imaging can be combined with optical coherence tomography (OCT) to enhance the resolution of superficial tissues, while plane wave imaging can integrate blood flow vector analysis to improve functional assessment capabilities. Third, clinical verification remains a necessary step, especially for the applications of endoscopic imaging and plane wave technology, which require large-scale testing in actual lesion scenarios to ensure stability, robustness, and diagnostic reliability.

% \begin{figure*}[h]
% 	\centering
% 		\includegraphics[scale=1]{elsevier-cas-double_column/wang_ruihua/f1.png}
% 	  \caption{Cardio axis}\label{FIG:2}
% \end{figure*}

\section{Conclusion}

This paper focuses on different application scenarios of ultrasound imaging, combining simulation and real data, and systematically completes the development and verification of the whole process of three typical imaging modes.
For line array imaging, a simulation system with 32 and 64 arrays is constructed based on the Field II platform, and the delayed-summing (DAS) beam synthesis algorithm is used for image reconstruction. The experimental results show that the increase in the number of arrays significantly improves the imaging resolution, in which the 64-array system improves the lateral resolution by about 40\% compared with the 32-array system.For endoscopic imaging, two simulation methods are proposed: one is a virtual ray-tracing method based on polar coordinates, which is suitable for rapid demonstration and structural validation; and the other is a ring-scanning simulation method based on the rotation of line arrays, which is closer to the actual physical mechanism. Both methods have successfully realized the stitching and visualization of intracavity images, which verifies the effectiveness of the methods.For plane wave imaging, multi-angle synthetic imaging experiments were carried out based on the publicly available PICMUS dataset to comprehensively analyze the influence of the number of excitation angles on the imaging quality. The results show that the synthesis scheme from single angle (C1) to 75 angles (C75) verifies the significant advantages of angle compositing in terms of image equalization and noise suppression.In summary, this project completes the complete technical link from modeling, simulation to image reconstruction, which lays a good foundation for subsequent algorithm optimization and system design.

% Use the style of numbering in square brackets.
% If nothing is used, default style will be taken.
%\begin{enumerate}[a)]
%\item 
%\item 
%\item 
%\end{enumerate}  

%\begin{itemize}
%\item 
%\item 
%\item 
%\end{itemize}  

%\begin{description}
%\item[]
%\item[] 
%\item[] 
%\end{description}  

%Figure
% \begin{figure}[h]
% 	\centering
% 		\includegraphics[scale=1]{elsevier-cas-double_column/wang_ruihua/f1.png}
% 	  \caption{Cardio axis}\label{fig:2}
% \end{figure}

% \begin{table}[h]
% \caption{Comparison of mechanical properties}\label{tbl1}
% \begin{tabular*}{\tblwidth}{@{}LL@{}}
% \toprule
%  Against Copper Film Test & Against Human Skin Test \\ % Table header row
% \midrule
%  Wet (10$\Omega$) & Wet (550$\Omega$) \\
%  Microneedle (1$\Omega$) & Microneedle (600$\Omega$) \\
%  Dry (0.1$\Omega$) & Dry (700$\Omega$) \\
% \bottomrule
% \end{tabular*}
% \end{table}

% Uncomment and use as the case may be
%\begin{theorem} 
%\end{theorem}

% Uncomment and use as the case may be
%\begin{lemma} 
%\end{lemma}



\bibliographystyle{cas-model2-names}

\bibliography{cas-refs}

% % Here goes the biography details.

% % Here goes the biography details.

\end{document}